\documentclass[letterpaper]{article} 
\usepackage{aaai2026}  
\usepackage{times}  
\usepackage{helvet}  
\usepackage{courier}  
\usepackage[hyphens]{url}  
\usepackage{graphicx} 
\usepackage{natbib}  
\usepackage{caption} 
\usepackage{algorithm}
\usepackage{algorithmic}

\usepackage{newfloat}
\usepackage{listings}
\DeclareCaptionStyle{ruled}{labelfont=normalfont,labelsep=colon,strut=off} 
\floatstyle{ruled}
\newfloat{listing}{tb}{lst}{}
\floatname{listing}{Listing}
\usepackage[table]{xcolor}
\usepackage{booktabs}

\title{Learning by Chatting? Investigating the Impact of Generative AI on \\Information Seeking and Learning}
\author {
    Shravika Mittal\textsuperscript{\rm 1,*},
    Su Lin Blodgett\textsuperscript{\rm 2,*},
    Q. Vera Liao\textsuperscript{\rm 3,*}
}
\affiliations {
    \textsuperscript{\rm 1}Georgia Institute of Technology\\
    \textsuperscript{\rm 2}Mila - Qu\'{e}bec AI Institute\\
    \textsuperscript{\rm 3}University of Michigan, Ann Arbor\\
    smittal87@gatech.edu, sulin.blodgett@mila.quebec,
    veraliao@umich.edu
}

\usepackage{bibentry}

\begin{document}

\maketitle

\renewcommand{\thefootnote}{*}
\footnotetext{Work done while at Microsoft Research Montr\'{e}al.}
\setcounter{footnote}{0}
\renewcommand{\thefootnote}{\arabic{footnote}}

\begin{abstract}
Generative AI (GenAI) tools offer increasing opportunities for augmenting human cognitive tasks. Among these tasks, information seeking is being rapidly reshaped by GenAI tools, with potentially profound implications for learning and knowledge acquisition. To investigate these implications, we conducted a between-subjects field experiment in which participants pursued informal learning by seeking information through either ChatGPT or Google Search over a span of 8 days. Using a daily diary protocol, we gathered in-situ data on their information-seeking processes. Our findings show that participants in the ChatGPT group experienced diminished agency in their information-seeking processes, as they offloaded much of the information selection to AI, and consequently experienced greater meta-cognitive load arising from this reduced sense of control. We further highlight two sources of distortion in information access when using chat-based GenAI tools: biases in ChatGPT outputs, particularly towards providing solution-oriented artifacts over principled knowledge; and systematic shifts in users' information-seeking behaviors, whereby the conversational and socially-oriented interaction paradigm of current GenAI tools may inadvertently reduce exploration of the broader knowledge space. As a result, on average, participants in the ChatGPT group had worse learning outcomes than those using Google, especially for higher-order critical learning. Our work suggests inherent tensions between offloading information seeking to AI and meaningful learning, and provides broader implications for understanding AI's risks to human cognition.
\end{abstract}

\section{Introduction}

In recent years, advances in Generative AI (GenAI) have shifted its role from primarily automating routine tasks to increasingly augmenting human capabilities~\cite{wu2025human}. This shift is evident in its widespread adoption across domains of knowledge work (e.g., software engineering)~\cite{ suri2024usegenerativesearchengines}, creative practices~\cite{suri2024usegenerativesearchengines}, and formal education~\cite{Nugroho04032025}. While GenAI tools can substantially improve productivity and efficiency, these gains are often realized through cognitive offloading~\cite{aiskillformation2026}---the practice of shifting effortful cognitive tasks to external aids. 

Among the various cognitive tasks, information seeking is one that has undergone a profound shift with GenAI-powered technologies. As users increasingly move away from the previously dominant paradigm of retrieval-based search to ``ask ChatGPT,'' they are offloading much of the mental labor involved in information seeking to chatbots and answer engines~\cite{10.1145/3498366.3505816}. In particular, GenAI tools offer advanced information synthesis capabilities (such as summary generation), combined with natural language understanding capabilities, to help process flexible and highly specified queries. While these affordances can support rapid information look-up, they also offload critical cognitive processes, such as selecting, evaluating, and verifying information, to AI systems~\cite{hirvonen2024artificial}. 

Information seeking is foundational to many downstream cognitive tasks such as learning, decision making, problem solving, and creativity~\cite{10.1145/3498366.3505816,spatharioti2025effects}. Consequently, shifts in how users seek information through GenAI tools may influence subsequent cognitive outcomes. In particular, emerging evidence in education suggests that unconstrained use of popular GenAI tools such as ChatGPT can impair learning. For example, in classroom settings, heavier use of ChatGPT has been associated with poorer academic performance \cite{bastani2024generative}, as GenAI tools can automate or accelerate answer-finding and task completion at the expense of deep conceptual understanding and genuine learning~\cite{10.1145/3706599.3719841}. 

Unlike prior work focused on formal educational settings, we examine GenAI's impact on \textit{informal learning}, through which individuals acquire knowledge and skills in everyday contexts~\cite{10.1145/3711062}. Whether to develop an artistic skill, learn a new language, or acquire practical life skills, individuals often rely on informal, self-directed pursuits mediated by information systems. While these learning pathways were once defined by search engines---studied under the paradigm of ``searching as learning''~\cite{10.1145/3176349.3176386}---and social media platforms~\cite{sengupta2020learning}, they are increasingly being reshaped by GenAI tools~\cite{terzimehic2025conversational,lira2025learning}. Crucially, informal contexts lack the institutional guardrails and instructional oversight found in formal settings, making users more likely to offload cognitive efforts that may be essential for building durable expertise.

To understand the mechanisms through which GenAI tools (and their unique affordances) may impact knowledge acquisition, we investigate the shifts in people's information-seeking behaviors during informal learning in naturalistic settings. We conducted a between-subjects, longitudinal field experiment (8 days) in which participants were randomly assigned to perform an informal learning task using either ChatGPT, a widely used GenAI tool~\cite{openaiIntroducingChatGPT}, or the conventional Google Search. In addition to measuring participants' knowledge gain,  we utilized a daily diary protocol to gather in-situ data about their information-seeking processes, and contrasted those for the two groups (ChatGPT vs. Google) through the lens of an information-seeking stage model~\cite{ellis1993comparison}. Our work seeks to answer the following research questions:  

\begin{description}
    \item [RQ1:] How does the use of GenAI tools impact people's informal learning outcomes? 
    \item [RQ2:] How does the use of GenAI tools impact people's information-seeking behaviors for learning?  
\end{description}

Our research demonstrates the risks of offloading information-seeking tasks to GenAI tools. With GenAI acting as a ``filter'' of knowledge spaces and providing fewer affordances for differentiating and assessing information, participants had diminished agency and control in their information-seeking processes, and experienced greater meta-cognitive load as a result of this diminished sense of control. In addition, we highlight two sources of \textit{distortion} in information access. First, distortion may arise from biases within the technology itself: ChatGPT currently exhibits biases in output generation---towards generating solution-oriented artifacts over principled knowledge, while providing little transparency or controllability over these biases. Second, distortion can occur via systematic shifts in people's information-seeking behaviors, in which the socially-oriented interaction paradigm of current GenAI tools, characterized by a ``chat'' interface and personalization as popular design choices, may inadvertently narrow query scope and reduce exploration of the broader knowledge space. This diminished agency and distortion in information-seeking processes were associated with worse informal learning outcomes, on average, in the ChatGPT group, particularly for higher-order critical learning. 


\section{Background and Related Work}

\subsection{Generative AI and Cognitive Offloading}
Cognitive offloading refers to the use of physical actions or external tools to reduce the mental demands of a task~\cite{risko2016cognitive}. People routinely offload cognitive efforts to digital systems in everyday contexts~\cite{risko2016cognitive}. Most notably, with the wide adoption of information retrieval systems such as Google Search, people increasingly offload information storage, remembering where to find information rather than the information itself~\cite{sparrow2011google}. Traditionally, such offloading has been understood as a way to redistribute cognitive effort, allowing individuals to outsource lower-level cognitive processes or sub-processes. Recent advances in GenAI introduce expanded forms of cognitive offloading. Unlike earlier digital tools that primarily support low-level, narrowly scoped cognitive tasks, GenAI tools increasingly support intermediate cognitive tasks such as synthesis, comprehension, and reasoning, and even automate complex cognitive workflows such as planning, decision making, and creativity. In other words, GenAI tools are increasingly integrated into human thinking processes, and in some cases, may even be replacing human thinking altogether~\cite{lee2025impact}. 

These shifts have spurred great public and academic interest in examining GenAI's implications for human cognition, especially in knowledge work~\cite{budzyn2025endoscopist} and educational settings~\cite{walker2025learning}. A series of studies has examined how the use of popular GenAI tools impacts specific cognitive tasks such as student learning~\cite{Yang03072025}, decision making~\cite{spatharioti2025effects}, creativity~\cite{kumar2025human}, and critical thinking~\cite{lee2025impact}. A few common observations have emerged from this literature. First, the use of GenAI tools involves cognitive offloading and may therefore lead to reductions in overall cognitive effort and engagement, including meta-cognitive activities~\cite{lee2025impact}. This is further evidenced by a study comparing EEG-based brain activity among participants using ChatGPT, a search engine, and no external tools~\cite{kosmyna2025your}.   

Second, this bypassing of cognitive effort impairs knowledge and skill acquisition~\cite{bastani2024generative,walker2025learning,10.1145/3706599.3719841}, contributes to deskilling~\cite{macnamara2024does}, and creates psychological harms such as loss of agency and ownership of one's work~\cite{kosmyna2025your}. Even if GenAI tools are found to enable performance boosts in some contexts, studies commonly show that knowledge workers and students do worse, compared to those not using AI from the beginning, when they lose access to these tools~\cite{kumar2025human}.


While situated in this broad research examining GenAI's impact on human cognition, particularly learning and knowledge acquisition, we identify a gap in the literature where the specific link between the increasing use of GenAI tools for information seeking and learning outcomes is under-investigated. As we review below, information seeking is a foundational component of learning, especially in everyday informal learning contexts where structured instruction and curricula are absent. Although GenAI offers greater opportunities for cognitive offloading than conventional retrieval-based search engines, especially through its information synthesis capabilities, information seeking remains a complex cognitive process whose transformation under GenAI warrants empirical examination.

\subsection{Information-Seeking Behaviors and Learning}\label{sec:rw_1}
In information science, a large body of work on ``searching as learning''~\cite{10.1145/3176349.3176386} shows that people use search engines not only to retrieve or look up information, but also to engage in learning processes such as sense-making~\cite{dervin1983overview} and bridging knowledge gaps~\cite{10.1145/3498366.3505816}. In fact, the specific behaviors or actions people adopt during their search support different cognitive modes of learning~\cite{doi:10.1177/0165551515615841,lee2015does}. For example, receptive learning---i.e., remembering and reproducing what is learned---is supported by behaviors like targeted fact retrieval. Critical learning---i.e.,  thinking differently with respect to one's own viewpoints---is supported by differentiating the value of information across multiple sources. Lastly, creative learning---i.e., generating new ideas or artifacts based on what is learned---is supported by sense-making. Therefore, scholars have looked at how search engines can be designed to better support search behaviors that lead to improved learning outcomes~\cite{allan2012frontiers,agosti2014evaluation,bates1989design}.

Many studies have produced models of information-seeking behaviors to elucidate how people retrieve information and to inform the design of better information systems~\cite{allam2019review}. Ellis's model of information-seeking behaviors~\cite{ellis1993comparison} identifies eight stages that people engage in when seeking information: starting, chaining, browsing, differentiating, monitoring, extracting, verifying, and ending. Wilson's model highlights that information-seeking behaviors are influenced by personal, social, and environmental factors~\cite{wilson1999models}, while Kuhlthau's model describes the interplay between affect and seeking behaviors~\cite{kuhlthau1991inside}. These models have been used to study seeking behaviors across various environments~\cite{ge2010information}. \citet{choo1999information} used Ellis's model as an analytical lens to study knowledge workers' information-seeking behaviors on the web.

Changes in information technologies are often followed by shifts in user behaviors~\cite{mitra2014user}, with GenAI representing the latest disruption~\cite{hirvonen2024artificial}. 
\citet{nngroupChangingSearch} found that the
``AI overviews'' panel appearing at the top of most search results captures users' attention and discourages them from visiting the actual web pages. \citet{10.1145/3491101.3519665} observed that participants prefer Copilot over search engines to initiate information seeking for programming tasks, but rely on human-vetted sources for information verification. 
Yet, we lack a systematic understanding of \textit{how} this new generation of GenAI tools reshapes information-seeking behaviors that are essential for learning.

To close this gap, we conducted a longitudinal field experiment with a diary study protocol comparing information-seeking behaviors using ChatGPT (a popular GenAI tool) and a conventional search engine. While researchers have begun to examine the differences between the two paradigms~\cite{spatharioti2025effects,10.1145/3613904.3642459}, prior work focuses on controlled lab settings, limiting their external validity, especially for learning-related behaviors and outcomes. The diary protocol allows us to collect in-situ data about how and why participants 
engage in all stages of information seeking, which we analyze through the lens of Ellis's stage model~\cite{ellis1993comparison}.

\section{Methods}

\subsection{Study Design}
We conducted a between-subjects, IRB-approved, longitudinal field experiment with a daily diary study protocol. Participants were recruited from Prolific (Section ``Participants''), a crowdsourcing platform widely used for human-subjects experiments. Over a period of 8 days (excluding weekends), participants were asked to use the assigned technology to seek information to learn about ``nutrition and meal planning'' (i.e., to perform an informal learning task)---a popular topic for informal learning~\cite{terzimehic2025conversational}. They were also encouraged to use (but were not restricted to) the technology for their other usual day-to-day information needs. They were randomly placed into one of two conditions, representing the assigned technology to use:
\begin{itemize}
    \item ChatGPT: a popular GenAI-powered application, 
    with information support (``\textit{ask anything}'') as one of its primary functions.
    \item Google: the long-standing popular search engine, which currently controls over 90\% of the global search engine market~\cite{statcounterSearchEngine}. Google Search recently introduced ``AI Overviews''~\cite{search}, a feature that uses GenAI to provide a snapshot of key information related to users' queries. To minimize infusion of direct GenAI outputs, we asked participants to append ``-ai'' (verified via screenshot review) to every search query to temporarily remove the panel~\cite{howtogeekTurnOverview}. 
\end{itemize} 
By comparing ChatGPT with Google Search without AI overviews, we captured two distinct ends of the information-seeking spectrum, i.e., GenAI-driven vs. conventional approaches.
The field experiment included three components:
\begin{itemize}
    \item Onboarding; Pre-learning knowledge test: Participants consented to the study and completed (1) a demographics survey, (2) a pre-learning knowledge test to capture their baseline knowledge about the learning topic, and (3) example tasks to prepare for the daily diary study (see below). Refer to Appendix for the onboarding survey.
    \item Diary study: Following onboarding, each day for eight days, participants were asked to complete two tasks as part of the diary study: (1) \textit{Informal learning task}: Every morning, we provided a guiding prompt, asking participants to learn about a subtopic related to the informal learning topic (see below). 
    They were required to use the assigned technology for at least 10 minutes and could use it however they preferred. While completing the learning task, they were asked to upload screenshots for any number of interactions with the technology and comment on them; 
    (2) \textit{Diary entry task}: After completing the learning task, participants were required to submit a diary entry reflecting on their information-seeking activities for the day. They were asked to reflect on (a) how they sought information to complete the learning task and (b) at least one additional instance of information seeking.\footnote{For the scope of this paper, we only focused on participants' responses to the first diary entry prompt.} 

    \item Post-learning knowledge test: Post experiment, participants completed a knowledge test designed to assess their learning outcomes.
\end{itemize}

\paragraph{Informal Learning Task.}\label{sec:informal_learning_desc}
We selected an informal learning task centered on acquiring a practical life skill~\cite{10.1145/3711062}---specifically, to learn about ``nutrition and meal planning.'' This topic fits the criteria of informal learning as it is self-directed, potentially personally relevant, requires active information seeking, and allows people to construct knowledge in informal contexts. Selecting this topic also enabled us to create a knowledge test and measure participants' learning outcomes, suitable for answering RQ1.

To facilitate learning over the eight days, we created broad daily guiding prompts, outlined in the Appendix (Section ``Guiding Prompts''), in reference to beginner-friendly, highly-rated, and heavily registered relevant courses~\cite{udemyNutritionCertification,udemyNutritionMasterclass}. For example, over the course of the eight days, participants were progressively asked to learn about macronutrients, how to plan meals tailored to a specific fitness goal, and strategies for eating healthy when traveling.

\paragraph{Diary Study Protocol.} \label{sec:diary}
Diary studies provide an opportunity to collect data on participants' situated practices in naturalistic settings over an extended period of time, while minimizing recall bias, making them particularly suited for RQ2~\cite{bolger2003diary}.
After completing the learning task and before the end of the day, participants were asked to spend at least 20 minutes writing diary entries reflecting on their information-seeking activities. To design our diary entry protocol (Appendix), we drew on prior work~\cite{10.1145/2858036.2858278}. 
The protocol asked participants to answer two open-ended questions: to (1) elaborate on how they sought information to learn about the day's subtopic using the assigned technology; and (2) describe how they addressed at least one additional information need. Under each question, we added prompts to guide participants to describe how they used the technology, what they learned and how they arrived at it, any challenges they encountered, the steps taken to address them, and their overall experience seeking information. We took several measures to ensure good quality diary data, including daily review of submissions, participant feedback, and safeguards against AI-generated entries (Appendix Section ``Diary Study: Quality Control'').


\subsection{Quantitative Measurement: Learning Outcome Score (RQ1)} \label{sec:learning_measurement}

\paragraph{Knowledge Test.}
We operationalized Bloom's taxonomy~\cite{anderson2001taxonomy} to design a knowledge test (Appendix Section ``Knowledge Test'') that measured participants' different levels of learning on the assigned topic following the eight-day study  (see ``Background and Related Work'' for the learning levels). Bloom's taxonomy is widely used in education to characterize different levels of learning objectives by cognitive complexity (from low to high): remembering, understanding, applying, evaluating and creating. To measure participants' \textit{receptive learning outcomes}, we gathered (with minor modification) multiple-choice questions from relevant courses on Udemy that focused on remembering, recalling, and understanding key concepts. 
To measure \textit{critical learning outcomes}, we designed a mix of multiple-choice and open-ended questions, asking participants to apply, analyze, or evaluate certain concepts based on what they had learned. For instance, participants were asked to 
critically evaluate two meal plans according to their nutritional quality. Relevant course materials, e.g., meal plans and end of course quizzes, from Udemy were used to create these questions. Finally, to measure \textit{creative learning outcomes}, we asked participants to create an artifact: in our case, a three-day meal plan.

We iteratively refined this knowledge test using feedback and responses from 11 pilot participants. Adjustments included removing questions that were answered correctly by a majority of the participants using just their baseline knowledge, and making minor tweaks to reduce ambiguity in the open-ended questions. We also ensured that obtaining the information needed to answer the questions required effort in both conditions (e.g., it could not be found in the first sentences or titles displayed when querying ChatGPT or Google Search with a relevant subtopic).

\paragraph{Scoring the Knowledge Test.}
The knowledge test consisted of nine multiple-choice and four open-ended questions. Since the multiple-choice questions were directly taken from courses on Udemy, we were able to score participants' responses against the course-provided ground truth using a binary correct (1), incorrect (0) scale. 

Through multiple discussions and in reference to existing guidelines on nutrition~\cite{mayoclinicHealthyMeals,usda_guidelines}, we iteratively developed rubrics to score participants' responses to the four open-ended questions, by first independently rating responses from seven participants in each group, and then discussing and resolving disagreements. Responses to two of the questions (Q11, Q12) for evaluating artifacts (nutrition labels and meal plans) were scored on a binary scale.
Responses to the other two questions (Q13, Q14) for improving or creating a meal plan were scored on a 1--3 scale, with higher scores for better responses according to the designed rubric (Appendix Section ``Rubric for Open-ended Questions''). 
Participants' scores on all knowledge test questions were summed to obtain their \textit{learning outcome score}.


\subsection{Qualitative Analysis of Diary Entries (RQ2)} \label{sec:qual_analysis}

To answer RQ2, we analyzed participants' responses to the first diary entry prompt, where they described how they sought information to complete the daily informal learning task. 
Prior work has proposed a range of models to characterize people's information-seeking behaviors~\cite{ellis1993comparison,kuhlthau1991inside,wilson1999models}. Among these, Ellis's model offers a particularly suitable analytical lens for our study, as while other models emphasize contextual 
or affective factors 
governing information seeking, Ellis's model focuses on observable behaviors, i.e., how information seekers interact in an information environment.

Ellis's model proposes eight behaviors people may engage in during their information-seeking processes: starting (\textit{initiating information seeking}), chaining (\textit{following connections to pursue a line of inquiry}), browsing (\textit{casually seeking information in potential areas of interest}), differentiating (\textit{using known differences to filter information}), monitoring (\textit{regularly following particular sources}), extracting (\textit{selectively identifying relevant information}), verifying (\textit{checking the correctness of information}), and ending (\textit{wrapping up}). It has been adopted to study behaviors of different user populations within varied information environments, both digital and offline~~\cite{choo1999information,ganie2019information,10.1145/3657604.3664657}. 
Using Ellis's model as a guiding framework, the first author conducted a preliminary thematic analysis of a subset of diary entries. Although the analysis began with a deductive coding-based approach, several sub-codes were generated inductively to capture participants' actions at a more granular level. These sub-codes further emerged as a consequence of the distinct ways in which participants engaged with ChatGPT or Google Search. Following this, the first author met with the other co-authors to discuss, refine, and finalize the coding framework. The final framework is summarized in Table~\ref{tab:qual_analysis}. 

After finalizing the codes and their definitions, the first author conducted a thematic analysis of participants' entries from days 2 to 8. Entries from day 1 were intentionally excluded, as participants were still familiarizing themselves with the writing process. In total, 98 entries from the ChatGPT condition and 140 entries from the Google condition were analyzed (with 7 entries missing from the Google condition). The difference in the number of entries across groups is due to the differing number of participants who completed the study in each group (see below).

\subsection{Participants}\label{sec:participants}

\paragraph{Recruitment.}
In total, we invited 144 participants on Prolific to complete a preliminary screening survey. Among those that fit our participation criteria (below), 80 were invited to participate in the experiment, recruited in two batches spaced one week apart (N1=50; N2=30). The second batch was recruited to account for participant attrition in the first batch (see below). 

The screening survey identified participants who (1) were moderate users of GenAI, (2) had moderate knowledge about and interest in ``nutrition and meal planning'' (the informal learning task), and (3) were likely to engage in informal learning. These criteria were intended to filter potential outlier behaviors, avoid floor and ceiling effects when measuring learning outcomes, ensure that participants engaged in the informal learning task voluntarily, and support participant retention. We also disclosed the topical focus of the experiment and screened out participants who might not feel comfortable seeking nutrition-related information.
Participants were compensated up to $\$105$; $\$8$ per day over a 10-day period for completing 30-minute daily tasks, and $\$25$ for a 45-minute semi-structured interview. Compensation was provided on a daily basis.

Of the 40 participants in each condition, 14 in the ChatGPT condition and 21 in the Google condition completed the entire study. The high attrition rate can be attributed to the nature of a diary study~\cite{ohly2010diary}. While not statistically significantly different, the higher drop-out rate in the ChatGPT condition is consistent with our results that participants in this group had more negative experiences and higher cognitive load. 
Participants who voluntarily dropped out expressed difficulty with writing diary reflections, uploading screenshots, or sustaining participation over the full study period. To maintain data quality, we also removed participants who consistently submitted low-quality diary entries despite receiving repeated feedback. We excluded data from participants who dropped out or were removed early.

\paragraph{Background.}~\label{sec:participant_background}
During recruitment, we aimed to achieve a balanced demographic distribution across the two groups. Among participants in the ChatGPT and Google conditions, respectively, 42.86\% and 19.05\% identified as women, 57.14\% and 76.19\% identified as men, and 4.76\% in the Google condition identified as non-binary. Age distributions were: 26--35 (21.43\% and 23.81\%), 36--45 (50.00\% and 42.86\%), 46--55 (21.43\% and 23.81\%), and above 55 (7.14\% and 9.52\%) for the ChatGPT and Google conditions, respectively. University or college was the most commonly reported level of education (42.86\% and 71.43\%), followed by high school (28.56\% and 23.81\%), and graduate degree (14.29\% and 4.76\%). 


During onboarding, participants completed a pre-learning knowledge test. This consisted of three multiple-choice questions (scores ranged from 0–3) from a subset of the final knowledge test (Q4, Q5, and Q8). We included only a subset of questions to keep the onboarding process at a reasonable length and to minimize potential recall effects during the final post-learning assessment. A Welch's t-test revealed no significant differences in the pre-learning topical knowledge between the two groups ($t$-statistic: $0.647$, $p$-value ($p$): $0.261$). On average, participants in the ChatGPT group scored $1.64$ (standard deviation (SD): $0.72$), and participants in the Google condition scored $1.48$ (SD: $0.73$).


\section{Results}

\subsection{Impact of GenAI on Learning (RQ1)}

\paragraph{Learning Outcome Score.} Given that the two groups started with comparable baseline knowledge on the assigned topic, we performed a Welch's t-test to compare participants' learning outcome scores as measured by the post-learning knowledge test. On average, participants in the ChatGPT and Google conditions scored $11.62$ (SD: $2.32$) and $12.86$ (SD: $2.21$), respectively. The test did not reach statistical significance but indicates a trend that participants in the ChatGPT condition had worse learning outcomes than those in the Google condition ($t$-statistic: $-1.523$, $p$: $0.069$). Regression analyses, controlling for baseline knowledge, showed a similar trend.


\subsubsection{Question-level Analysis}
Next, since the questions were designed to cover different aspects (i.e., subtopics and levels of learning), we performed an exploratory, question-level analysis to examine on which aspects participants in the two groups showed different learning effects. Specifically, for each multi-choice question, we performed a Fisher's exact test, recommended for comparing binary outcomes within small samples, on participants' answer correctness between the ChatGPT and Google conditions. This analysis revealed that participants in the Google group were significantly more likely to correctly answer two of the multiple-choice questions designed to assess critical learning, i.e., application or evaluation of concepts learned.

First, one of these questions (Q8) required participants to apply their knowledge and interpret a given nutrition label. In the post-learning knowledge test, a significantly smaller proportion of participants in the ChatGPT condition answered Q8 correctly, compared to those in the Google group ($57\%$ vs. $95\%$; odds ratio: $0.07$; Fisher's $p$: $0.01$). This difference was not significant when the participants attempted the same question in the pre-learning knowledge test ($64\%$ vs. $62\%$; odds ratio: $0.56$; Fisher's $p$: $0.47$). Based on qualitative analysis of participants' diary entries on the day they were assigned a subtopic relevant to Q8 (day 6), the ChatGPT group's worse performance could be attributed to their more limited exposure to realistic and diverse examples of nutrition labels. For example, one participant noted (P6; ChatGPT), \textit{``while learning with ChatGPT, one aspect I found unsatisfying was the lack of visual examples, which made it harder to apply the advice directly,''} while another reflected (P7; ChatGPT), \textit{``I wish [ChatGPT] could give me an example of a real nutrition label to gain more practical knowledge.''} This limitation forced some participants to engage in additional information-seeking activities from other channels: \textit{``I paired the guidance with labels from products in my pantry''} and \textit{``I will now try [to get examples of some real nutrition labels] using Google.''}  

Next, the other question (Q7) required participants to compare and evaluate animal and plant sources of protein. Again, a significantly smaller proportion of participants in the ChatGPT condition answered correctly compared to those in the Google condition ($21\%$ vs. $57\%$; odds ratio: $0.21$; Fisher's $p$: $0.04$). This difference could stem from content bias within ChatGPT: in the diary entries, participants reflected that ChatGPT by default provided meat-based sources of protein. For example, one participant (P3; ChatGPT) noted \textit{``[ChatGPT] automatically assumes that everyone eats meat.''} Taken together, these limited and biased outputs provided by ChatGPT, combined with participants' reduced exploration and agency in learning, as discussed in the qualitative results below, may have contributed to their worse learning outcomes, particularly those for higher-level critical learning.

\subsection{Impact of GenAI on Information-Seeking Behaviors (RQ2)}

As described earlier, we conducted a thematic analysis of participants' diary entries using the coding framework summarized in Table~\ref{tab:qual_analysis}. $n_C$/$n_G$ represents the number of entries from participants in the ChatGPT/Google condition that mentioned the information-seeking behavior or code. These numbers should not be interpreted as quantitative measurements, but serve to guide our attention towards how information-seeking behaviors differed in the two conditions. We present the main contrasts below.

\begin{table*}[!ht]
    \centering
    \small{
    \begin{tabular}
{@{}p{0.25\textwidth}|p{0.45\textwidth}|r|r|r|r@{}}
    \toprule
        \textbf{Behavior (or Code)} & \textbf{Description} & \textbf{$n_C$} & \textbf{$n_G$} & $\%_C$ & $\%_G$ \\
    \midrule
        \rowcolor{black!10} Starting (broad) & Beginning with exploratory, wide-ranging queries to serve as starting points for the information-seeking process. & $66$ & $135$ & $67.35$ & $96.43$ \\ \hline
        \rowcolor{black!10} Starting (narrow) & Beginning with specified, focused queries (e.g., tailored to personal preferences or well-defined needs) to serve as starting points for the information-seeking process. & $32$ & $5$ & {$32.65$} & {$3.57$} \\ \hline
        Chaining (self-initiated) & Following a line of inquiry by using follow-up queries, links, or citations that build on previously retrieved information. & $36$ & $39$ & {$36.73$} & {$27.86$} \\ \hline
        Chaining (technology-mediated) & Following a line of inquiry by using queries, links, or citations suggested by the technology (e.g., via ChatGPT-suggested follow-up prompts, a previous interaction, or ` people also search for'' panel in Google). & $4$ & $7$ & {4.08} & {5.00} \\ \hline
        \rowcolor{black!10} Browsing & Casually seeking information in potential areas of interest, without a specific information need. & $9$ & $62$ & {$9.18$} & {$44.29$}\\ \hline
        Differentiating (based on personal relevance) & Using indicators of personal relevance to filter information. & $43$ & $68$ & {$43.88$} & {$48.57$} \\ \hline
        Differentiating (based on credibility) & Using indicators of credibility to filter information. & $0$ & $43$ & {$0$} & {$30.71$} \\ \hline
        Differentiating (based on type of information need) & Using indicators of the type of information needed (e.g., level of detail, format, topic) to filter information. & $0$ & $14$ & {$0$} & {$10.00$} \\ \hline
        \rowcolor{black!10} Monitoring & Regularly following particular sources of information (e.g., a trusted source, or a source encountered earlier). & $0$ & $26$ & {$0$} & {$18.57$} \\ \hline
        Extracting (information) & Selectively identifying relevant fundamental or principle-oriented information. & $65$ & $136$ & {$66.33$} & {$97.14$} \\ \hline
        Extracting (artifacts) & Selectively identifying artifact-oriented information, such as meal plans, grocery lists, or nutrient-intake trackers. & $46$ & $9$ & {$46.94$} & {$6.43$} \\ \hline
        \rowcolor{black!10} Expressing an intent to verify & Indicating an intention to verify information, without actually performing the verification action. & $3$ & $0$ & {$3.06$} & {$0$} \\ \hline
        \rowcolor{black!10} Verifying (via another source) & Checking the correctness of information by consulting another source, such as a different information channel or another webpage. & $3$ & $27$ & {$3.06$} & {$19.29$} \\ \hline
        \rowcolor{black!10} Verifying (via prior knowledge) & Checking the correctness of information via prior knowledge. & $4$ & $4$ & {$4.08$} & {$2.86$} \\ \hline
        Ending (Satisfied) & Concluding the information-seeking process with overall satisfaction (information need was met), even if minor frustrations or interruptions were expressed. & $86$ & $125$ & {$87.76$} & {$89.29$} \\ \hline
        Ending (Unsatisfied) & Concluding the information-seeking process with clear dissatisfaction, as the information need remained unmet. & $12$ & $15$ & {$12.24$} & {$10.71$} \\
    \bottomrule
    \end{tabular}}
    \caption{Summary of the codebook used to study participants' information-seeking behaviors as described in their diary entries. $n_C$/$n_G$ represents the number of entries from ChatGPT/Google participants. {$\%_C$/$\%_G$ represents the percentage of total entries from each group that described a given behavior.} We analyzed $98$/$140$ diary entries from the ChatGPT/Google groups.}
    \label{tab:qual_analysis}
\end{table*}

\paragraph{Reduced Querying Scope in the ChatGPT Condition.} \label{sec:reduced_exploration}
The frequencies of the ``Starting'' codes (see Table~\ref{tab:qual_analysis}) show that participants in the ChatGPT condition were more likely to begin their information-seeking with a narrow or specified query, a behavior observed in $32.65\%$ of their entries ($n_C$=32/98) compared to only $3.57\%$ of entries from Google participants ($n_G$=5/140). This suggests that interacting with ChatGPT may have constrained participants' information-seeking trajectories and reduced their exploration of the knowledge space from the very beginning. 

The narrower start happened in two ways: First, instead of seeking a general overview of the informal learning topic, participants tailored their conversational queries to get specific, personally relevant information (see excerpt below):

\begin{quote}
    ``I used ChatGPT to learn about macronutrients, specifically, by asking how much of each a person should consume per day by providing a rough estimate of my height, weight and activity levels.'' -- ChatGPT participant (P3)
\end{quote}

This behavior could be attributed to the conversational interface invoking a social interaction-oriented schema~\cite{lee2010receptionist}: \textit{``[feels] as if I am having a conversation''} and \textit{``led to an interesting conversation.''} In some cases, participants even referred to it as a nutritional coach: \textit{``It felt as though I was getting a nutritional coach for free.''} In contrast, participants in the Google condition started their information-seeking process with a broader query: one participant noted, \textit{``I searched how much of each macronutrient is recommended for a balanced meal.''} 

Second, ChatGPT's interface affordance---a chat-based UI enabling easy revisiting of prior interactions---made participants more inclined to continue their previous interactions rather than initiating a completely new query:

\begin{quote}
    ``To learn about today's topic, I simply opened my prior interaction with ChatGPT and asked it to summarize all the lessons learned so far.'' -- ChatGPT participant (P3)
\end{quote}

Moreover, even when participants began with broad queries, ChatGPT’s context retention and personalization limited exploration:

\begin{quote}
    ``I asked ChatGPT how much of each macronutrient one should have. Since my information was saved from yesterday, it gave me a personalized guide.'' -- ChatGPT participant (P11)
\end{quote}

\noindent In short, certain affordances of ChatGPT, including conversational interactions and easy access to and personalization based on past interactions, can constrain what people choose to query~\cite{10.1145/3498366.3505816}, reducing knowledge exploration. This echoes prior findings that conversational interactions can bias people towards issuing personal, opinionated, and highly specified inputs, potentially reinforcing ``information bubbles'' ~\cite{10.1145/3613904.3642459}.

\paragraph{Reduced Information Differentiation and Verification in the ChatGPT Condition.} Participants in the ChatGPT condition reported fewer instances of differentiating information, i.e., assessing (and filtering) information based on specific criteria (e.g., personal relevance, credibility, type of information need) and their indicators (``Differentiating'' in Table~\ref{tab:qual_analysis}, $43$/$125$ instances in ChatGPT/Google conditions). 

First of all, they applied a narrower set of criteria to filter information, relying almost entirely on personal relevance:

\begin{quote}
    ``I no longer eat meat, so [...] I focused on the vegetarian foods that were recommended by ChatGPT.'' -- ChatGPT participant (P12)
\end{quote}

On the other hand, participants in the Google condition considered additional indicators, such as source credibility and the type of information need to filter: 

\begin{quote}
    ``I looked for sources that I know to be reputable [...] I clicked on an NIH result.'' -- Google participant (P7)
\end{quote}


These differences in information differentiation can be attributed to offloading the information selection process to ChatGPT's generative capabilities, and ChatGPT providing fewer affordances for people to make their own selection. While intended to provide fast and convenient information access, this may deprive information seekers of control over their own information-seeking processes. For example, some participants in the ChatGPT condition expressed a reduced sense of control, and noted that they were unable to learn in the ways they would normally prefer:

\begin{quote}
    ``I prefer to seek out my own sources. Using [ChatGPT] is just not what I would prefer. [When learning] I like to read as much or as little detail as I would like.'' -- ChatGPT participant (P2)
\end{quote}

In contrast, participants in the Google condition noted that the ability to differentiate information sources based on their information needs gave them greater control over what to consume. For instance, one participant highlighted the ability to switch sources depending on whether they sought general advice or more nuanced, scientific information: 

\begin{quote}
    ``I used a Reddit thread and a blog, which worked well for tips and advice, but earlier in the week, when I was searching for information on macronutrients and scientific topics, these sources were less helpful. I  was able to adjust my sources accordingly as my information needs changed.'' -- Google participant (P7)
\end{quote}

Next, in the verifying stage (after information extraction), there were only seven instances where ChatGPT participants checked the accuracy of the information. They either did so by consulting another source (three instances, two by the same participant), such as referring to official health-related web pages (e.g., \textit{``cross-referenced some of the suggestions with official health websites''}) or by relying on their own prior knowledge (four instances; e.g., \textit{``I have also heard that mentioned elsewhere''}). Interestingly, a few reflections mentioned an \emph{intent to verify} the information obtained (three instances); however, they did not describe engaging in the actual verification behavior. Such an intent was expressed as skepticism about how ChatGPT worked (e.g., \textit{``I am not sure how accurate that is since the data is probably from 2023''}) and whether the information provided by it could be trusted (e.g., \textit{``[seeking health-related information] is not something I [would] trust a chatbot with right now''}).



Meanwhile, as reflected in the frequency of the ``Verifying'' code (Table~\ref{tab:qual_analysis}), participants in the Google condition verified information more frequently ($31$ instances). Participants often did so by consulting multiple web pages:

\begin{quote}
    ``I relied on an article from a website I considered reliable. It all looked legitimate, but I fact-checked to be sure. [...] I looked at several other websites quickly.'' -- Google participant (P6)
\end{quote}

Overall, these findings suggest a shift toward a more passive form of learning~\cite{bonwell1991active} in the ChatGPT condition, as ChatGPT provides synthesized information in its output at the cost of taking away information seekers' agency in filtering, assessing, and selecting information. This not only results in diminished sense of control over one's learning process, but can also be especially detrimental to critical and creative learning outcomes~\cite{doi:10.1177/0165551515615841}.


\paragraph{ChatGPT Reduced Extraction of Fundamental or Principle-Oriented Information.} As shown by the frequencies of the ``Extracting'' codes in Table~\ref{tab:qual_analysis}, participants in the ChatGPT condition were less likely to extract fundamental or principle-oriented information (i.e. the underlying principles for how things should be done and why, such as information about the recommended amounts of macronutrients) as takeaways from their information-seeking process. $66.33\%$ ($n_C$=65/98) entries from ChatGPT participants described this behavior, compared to $97.14\%$ ($n_G$=136/140) of entries from participants in the Google group. They were more likely to extract solution-oriented, artifact-based information, such as meal plans, grocery lists, or nutrient-intake PDF trackers, compared to the Google group (``Extracting (artifacts)'' in Table~\ref{tab:qual_analysis}); this behavior was observed in $46.94\%$ ($n_C$=46/98) of entries from ChatGPT participants, compared to only $6.43\%$ ($n_G$=9/140) of entries from the Google group.

Even when a participant queried to obtain principle-oriented information, such as recommended macronutrient amounts for a balanced meal, ChatGPT provided a sample meal plan, followed by a suggested grocery list:

\begin{quote}
    ``I asked ChatGPT how much of each macronutrient one should have for a balanced meal. It gave me a balanced meal plan. [...] It asked if I'd want a grocery list, which it provided.'' -- ChatGPT participant (P11)
\end{quote}

By contrast, a similar query by a participant in the Google group led them to learn about the principles of recommended amounts of macronutrients:

\begin{quote}
    ``I searched how much of macronutrients are needed for a balanced meal. A dietitian blog explained: carbs (45\% to 65\% of total calories), protein [...] and fats [...], for an average diet.'' -- Google participant (P6)
\end{quote}

Receiving solution- or artifact-oriented information may also lead participants to expect to rely on ChatGPT for situations where they need to apply the knowledge, reducing their motivation to develop the expertise in themselves~\cite{CRONINGOLOMB2023103816}, as shown by the quote from the same participant (P11): 

\begin{quote}
    ``I plan to use ChatGPT, even after this study, to create meal plans for me. It has always been overwhelming for me to do so.'' -- ChatGPT participant (P11)
\end{quote}

In short, ChatGPT currently exhibits a bias towards generating solution-oriented artifacts, rather than principle-based information to allow someone to build fundamental knowledge about how things should be done and why. This bias could potentially be a by-product of training ChatGPT as a product targeting knowledge workers who may rely heavily on artifact generation (e.g., programming assistance, writing support), making it a mismatch for learning purposes.


\paragraph{Unsatisfied Ending from Meta-cognitive Load in the ChatGPT Condition.} Participants in the ChatGPT condition were slightly more likely to end their information-seeking process expressing dissatisfaction (``Ending (Unsatisfied)'' in Table~\ref{tab:qual_analysis}). The code was observed in $12.24\%$ of entries from participants in the ChatGPT group ($n_C=12/98$), compared to $10.71\%$ of entries from those in the Google group ($n_G=15/140$).

In the ChatGPT condition, many unsatisfactory endings were rooted in meta-cognitive load and frictions~\cite{tankelevitch2024metacognitive}---i.e., the extra mental effort required to monitor and attempt to control one's information-seeking process and the frustrations from failures in doing so:

\begin{quote}
    ``I asked: shouldn't a person consult a doctor before making any major changes. I was frustrated when it said it's good to consult a doctor but not necessary. [...] I don't agree with this. I got more direct and expressed my concerns [...], asking ChatGPT if it was okay for it to provide that advice. Again, it hedged to say that it tries to provide information that does not harm anyone. [...] Using ChatGPT was frustrating and I just left the chat.'' -- ChatGPT participant (P2)
\end{quote}

Due to a lack of affordances for controlling information selection from ChatGPT's outputs, the main path through which one can control their information seeking is prompting and re-prompting. However, prompting  remains an unpredictable task where users do not always get the intended output (the right content in the right format), referred to as the ``instruction gap''~\cite{subramonyam2024bridging}, as illustrated in the excerpt below:

\begin{quote}
    ``ChatGPT chose to provide two separate answers this time. It provided the information in a different layout, which I found  confusing. It made me second-guess my query.'' -- ChatGPT participant (P8)
\end{quote}

More often than not, this burden on users led them to question their own capability for using ChatGPT:

\begin{quote}
    ``Users have to learn how to use ChatGPT. [...] I tried my best to then ask a basic query: how to go about losing weight, but the assistant again gave me responses that included information on both losing weight AND gaining muscle.'' -- ChatGPT participant (P4)
\end{quote}



In some instances, the inability to obtain information in a desirable format led participants to abandon their information-seeking altogether:

\begin{quote}
    ``[ChatGPT] listed the macronutrient quantity in grams. I was confused, as I have no idea how to measure food in grams. [...] This was of no use to me.'' -- ChatGPT participant (P7)
\end{quote}



In the ChatGPT condition, even when information needs were met in the end (among those who expressed satisfaction), participants experienced heightened meta-cognitive load because suggestions or follow-up prompts from ChatGPT were distracting them from their learning process and they had to make a conscious effort to stay on track:

\begin{quote}
    ``ChatGPT's follow-up prompts were interruptive. They steered the conversation towards diet-related topics, even when I was asking more science-focused questions.'' -- ChatGPT participant (P6)
\end{quote}

\begin{quote}
    ``ChatGPT was `eager' to give me a sample meal plan, even though I was asking about something else altogether. I declined.'' -- ChatGPT participant (P11)
\end{quote}

These results further show that participants in the ChatGPT condition experienced a diminished sense of control in their information-seeking process. This placed high meta-cognitive demands and frictions on them to monitor, attempt to regain control over, or regulate their interactions to stay ``on track'' with their learning goals. Currently, ChatGPT provides few affordances for users to meet these demands, leading to dissatisfaction and abandonment of the system. 








\section{Discussion}
Through a between-subjects, longitudinal field experiment using a diary study protocol, we found evidence that using ChatGPT for informal learning may lead to worse learning outcomes than using conventional search engines, especially for higher-order critical learning~\cite{lee2015does}. Participants' diary entries revealed that the worse learning outcomes in the ChatGPT group can be attributed to their reduced exploration of the knowledge space and lack of agency in information selection, combined with limited and biased information provided by ChatGPT, leading to an overall more unsatisfactory learning experience from a diminished sense of control. Reflecting on these observations, below we discuss two implications of cognitive offloading to AI for information seeking, learning, and broader contexts.


\subsection{Fundamental Tensions between Cognitive Offloading and Learning}
Compared to conventional search engines, which arguably allow users to offload the cognitive tasks of information retrieval and storage, GenAI tools' synthesis capabilities can offload much of the extended processes of collecting, selecting and assessing information to AI, resulting in diminished human agency and control over one's own information-seeking process. While potentially bringing value via, e.g., improving the efficiency and potentially the quantity of information access, this shift of agency may present fundamental tensions for learning. As our qualitative results suggest, learners may lose the means and motivation to thoroughly explore the knowledge space, miss opportunities to engage in information filtering and assessment, which are key to critical learning and developing durable knowledge, and become passive receivers of information, forming only surface-level understanding with an ``illusion of knowledge''~\cite{walker2025learning}. Since agency and control play critical roles in learning, we also observe participants experiencing heightened meta-cognitive load and frustration, and overall unsatisfactory user experience with GenAI tools for learning purposes. 

Our results echo, and provide explanations for,  recent literature in the education field which finds that the ``default'' designs of GenAI tools, which are oriented towards cognitive offloading for productivity gain, combined with students' unconstrained use without guardrails, are often detrimental to learning~\cite{bastani2024generative,10.1145/3706599.3719841}. Notably, research efforts have emerged to develop dedicated, structured GenAI tools for education, which often focus on cognitive \textit{scaffolding} instead of offloading, such as providing feedback, prompting reflection, and engaging in Socratic questioning to facilitate structured exploration of the knowledge space~\cite{tabarsi2025herald}. However, the development and adoption barriers for these structured tools remain steep, especially as unstructured, ``general-purpose'' GenAI tools such as ChatGPT are easily accessible.

Outside formal education settings, people are even more likely to continue using these general-purpose GenAI tools for knowledge and skill acquisition, in both planned and unplanned fashions. To preserve and even enhance these opportunities for informal learning we suggest two areas to develop new sets of affordances in GenAI tools in addition to the scaffolding approaches explored in educational technology research. One area is to provide affordances for users to understand and control which processes or subprocesses---e.g., which of Ellis' model's information-seeking stages---are offloaded to AI. Such modular support would allow GenAI tools to be flexibly configured to meet different goals (e.g., learning versus productivity) and the demands of diverse cognitive tasks. Second, the (sometimes unbounded) opportunities to offload cognitive tasks to AI call for additional meta-cognition support~\cite{tankelevitch2024metacognitive}: improving users' awareness of their own goals, processes, capabilities, and---in learning contexts in particular---their self-regulation capabilities~\cite{ertmer1996expert} to take charge of their own learning goals while using AI; monitoring their progress; and managing their motivation, emotions, and behaviors, including their use of AI.

\subsection{GenAI's Risks in Cognitive Distortion}

GenAI tools are often framed as opportunities for intelligence augmentation~\cite{engelbart2023augmenting}, offloading some cognitive tasks to extend human capabilities. This framing often invokes the analogy of the introduction of calculators, which undoubtedly have extended people's mathematical capacities in everyday life.  However, unlike calculators with direct control of input and deterministic output, our results highlight two forms of distortion risk---systematic change in the outcome of a cognitive process as compared to without AI---associated with cognitive offloading to GenAI: (1) distortions introduced through the generated outputs themselves, and (2) distortions introduced through systematic shifts in people's interaction behaviors. Together, these distortions can reshape downstream cognitive tasks, learning processes, and broader patterns of information engagement.

First, biases are inevitable with GenAI outputs. The generative synthesis capability acts as ``information filters'' of a much broader information space, and these filters are not value-free, but laden with values and judgments of technologies' developers. For example, we observed that ChatGPT is currently biased towards generating solution-oriented artifacts rather than principle-based information. This is likely influenced by the technology developers' biases towards supporting artifact-oriented knowledge work tasks (e.g., coding or writing). As user logs of GenAI tools report heavy use towards knowledge work-related tasks \cite{suri2024usegenerativesearchengines,handa2025education}, these usage patterns or user bases likely feed back into these technologies' training and design decisions. Unfortunately, such a bias runs counter to the fundamental purpose of learning: to gain principled knowledge about \textit{how} and \textit{why}~\cite{lee2015does}. Problematically, these biases are often not understood, nor made transparent or controllable, leaving the distortion effects of cognitive loading potentially long-lasting but unexamined.

Second, our results suggest that this distortion risk can also result from systematic shifts in people's interaction behaviors, i.e. what and how one offloads to AI. These behavioral shifts can inadvertently result from seemingly popular design choices of GenAI tools. For example, echoing prior work~\cite{10.1145/3613904.3642459}, we found that the social, chat-based paradigm had unintended consequences on narrowing participants' information querying. Certain technical affordances that have been extolled for the new generation of ``companion''-like GenAI tools, such as context preservation, personalization, and proactive interactions, may risk further homogenizing people's interactions with these tools. We urge future research to more carefully examine how various design choices interfacing with people's offloading of different cognitive tasks can create systematic shifts from how these tasks are performed without AI.\looseness=-1

\subsection{Limitations and Future Work}
We note several limitations of our study. First, owing to our study design, the informal learning task was not completely self-directed or unstructured. Nevertheless, while we provided broad topical prompts to help initiate participants' learning, we emphasized that these prompts were only meant as guidance (see Appendix), and participants were free to explore and engage with the task in ways that felt natural and meaningful to them. Second, the extent to which people engage in cognitive offloading likely varies across individuals~\cite{OBRIEN2017244}, learning strategies, and patterns of GenAI use. We focus on aggregate behavioral differences rather than modeling  individual differences. Therefore, our findings should not be interpreted as characterizing all learners or all forms of learning. Future work should investigate the interplay between individual characteristics and information-seeking behaviors. Further, behaviors may also be influenced by the information domain. For example, in high-stakes contexts such as healthcare, information-seeking behaviors such as differentiation based on credibility or verification may be more pronounced~\cite{10.1145/506443.506508}. We used a diary study protocol to capture participants' in-situ seeking behaviors. We acknowledge that these accounts may represent only some of these behaviors, as participants may not have documented every instance or fully articulated the nuances of their information-seeking process. Next, through this work, we do not make any causal claims linking information-seeking behaviors to learning outcomes. We provide a speculative discussion describing how certain behaviors may have led to particular outcomes. Future work should investigate this relationship empirically. Lastly, participant attrition led to a small sample size. Nonetheless, our final sample remained within the recommended range for diary studies~\cite{nngroupDiaryStudies}.

\subsection{Conclusion}

People seek information not only to retrieve facts, but to acquire knowledge and develop expertise. GenAI-powered applications are increasingly mediating access to information, and, for many, replacing conventional search engines. This paradigm change involves offloading of the information-seeking process to AI. To understand how GenAI tools may reshape information seeking and impact learning, we conducted a between-subjects, longitudinal field experiment comparing participants using ChatGPT versus Google Search to pursue informal learning. We found that, by offloading much of the information selection, the ChatGPT group had diminished agency in their information seeking and took a more passive role in learning. This presents fundamental tensions with learning, which also manifested as heightened meta-cognitive demands and unsatisfactory user experience for the ChatGPT group in attempting to regain control in their learning processes. The diminished agency led to, on average, worse learning outcomes in the ChatGPT group, accompanied by two sources of distortion that systematically shifted their information access. First, ChatGPT exhibits a bias toward generating solution-oriented artifacts over principled knowledge about `how' and `why.' Second, the socially-oriented interaction paradigm of current GenAI tools may have unintended consequences in shifting people's information-seeking behaviors. Popular design choices such as the ``chat'' interface and personalization may inadvertently reduce exploration of the knowledge space. 

\section{Ethical Considerations Statement} The study design was approved by our organization's internal IRB and we obtained participant consent. Following guidelines provided by Prolific~\cite{prolificProlificapossPayment}, participants were compensated at an hourly wage of 15 USD. We paid and approved the work of everyone who completed the study, regardless of whether their responses passed our quality checks. We did not collect any personally identifiable information, except for participants' Prolific IDs for compensation and communication purposes. All communication between the research team and participants was conducted anonymously via Prolific's internal messaging system~\cite{prolificContactParticipants}. Participants were free to discontinue at any time without penalty. Participants were compensated on a daily basis for their contributions. All participant data were anonymized prior to analysis.

\section{Acknowledgments}
We thank our participants for their time and effort. We also thank the FATE groups at Microsoft Research Montr\'{e}al and NYC, Daniel Russell, and Jiawei Zhou for thoughtful discussions and feedback. 

\bibliography{aaai2026}

\appendix
\section{Appendix}

\section{Diary Entry Protocol} \label{sec:diary_protocol}

\subsection{Prompt 1}
To help us understand how you used the assigned technology to learn about our provided topic for the day, please describe in detail what you did, for example the types of questions or queries you tried, what you learned and how you arrived there, any parts of the learning process that felt unsatisfying (or challenging) and how you addressed them, and anything else you would like to share about your overall experience learning with the technology. Below is a sample paragraph you can use as a template, but feel free to write it in your own words.

To learn about the assigned topic, I [asked ChatGPT/used Google to search for] \rule{1cm}{0.15mm}. Some examples of useful things I learned include \rule{1cm}{0.15mm}. I learned these by asking [ChatGPT/Google] \rule{1cm}{0.15mm}. While learning with [ChatGPT/Google], one aspect I found unsatisfying or challenging was \rule{1cm}{0.15mm}. To address this, I \rule{1cm}{0.15mm}.  Overall, I felt the learning process using [ChatGPT/Google] was \rule{1cm}{0.15mm} because \rule{1cm}{0.15mm}.

\subsection{Prompt 2}
Besides the assigned topic, recall at least one other instance (even better if you can provide more) today where you sought information, whether by using [ChatGPT/Google] or other channels. Describe your information need (for what topic or question you needed to seek information), how you tried to satisfy this need (e.g., what sources, queries, or strategies you used, and why you chose them), what the outcome was, and how you felt about this information-seeking experience. Describe any parts of the information-seeking process that felt unsatisfying or challenging, and how you addressed them.

\section{Diary Study: Quality Control}
One of the authors reviewed participants' submitted diary entries to provide feedback---appreciating high-quality submissions and suggesting any improvements. We also ran a small-scale pilot to improve our protocol. Participants were explicitly instructed not to use any AI tools when writing diary entries, and the copy-paste functionality was disabled in the survey to prevent this. The first author conducted daily checks to ensure that the diary entries were reflective of what was asked for in the prompt and of sufficient quality. To further encourage participation, we sent friendly reminders towards the end of the day to submit the diaries. We allowed participants to skip submitting entries occasionally, i.e., up to two consecutive days, acknowledging that participants' everyday responsibilities could prevent them from daily participation~\cite{janssens2018qualitative}.

\section{Guiding Prompts} \label{sec:learning_prompts}

\begin{enumerate}
    \item [Day 1:] Hi! We hope that you are excited to start learning about nutrition and meal planning. As a reminder, the sub-topics we provide each day are simply guiding prompts. Feel free to explore and learn in a way that feels natural and meaningful to you. 

    Let’s get started! Before you can plan a balanced meal, it’s important to understand the basics of \textbf{macronutrients}, which define what a healthy meal should include. Take some time today to learn about the different types of macronutrients, the roles they play in the body, and examples of nutrient-dense sources for each. 

    \item [Day 2:] Now that you have gained a basic understanding of macronutrients, take some time to learn about \textbf{how much of each macronutrient is recommended} for a balanced meal. 
    Explore how these recommendations can vary depending on one’s fitness goals, demographics, and activity level. To make this more interesting to you, you may also seek recommendations for your own fitness goal, age, or activity level. 

    \item [Day 3:] We hope you now have a good understanding of what makes up a balanced meal. 
    Today, we invite you to explore another key aspect of healthy meal planning: \textbf{meal timing}. When and how often should one consume different macronutrients? There is no universal rule, so learn about different approaches to meal timing and how they may influence energy levels, metabolism, and overall well-being.  
    You may consider exploring a meal timing approach that fits your personal schedule and lifestyle. 

    \item [Day 4:] Let's bring together everything we’ve learned so far and \textbf{focus on a specific fitness goal: building muscle}. Take some time today to learn how to set up a meal plan that supports muscle growth, including key nutrients, meal timing, and calorie considerations. 

    Next, \textbf{focus on a different fitness goal: weight loss.} Learn how to set up a meal plan that supports gradual, sustainable weight loss (without muscle loss), including key nutrients, meal timing, and calorie considerations. 

    \item [Day 5:] It’s time to learn about \textbf{meal prepping}, i.e., practical strategies to turn meal plans into action. Meal prepping can help save time, reduce stress, and stay on track with one’s nutrition goals by ensuring that healthy, balanced meals are always within reach. Learn about different approaches to meal prepping, such as batch cooking, ingredient prep, meal portioning.  

    Also explore how grocery shopping plays a key role in successful meal prep: how planning ahead, making a clear shopping list, and choosing a variety of ingredients can set you up for an efficient, cost-effective week. 

    \item [Day 6:] Take some time to learn \textbf{how to read and interpret nutrition labels}, a key skill for making informed food choices that align with your meal plan and health goals. Explore how to evaluate the nutritional profile of a product using food labels and learn which components are most important to pay attention to, such as serving size, calorie content, added sugars, saturated fats, and fiber.  

    Also, learn how to use the ingredient list to identify products worth buying or avoiding. 

    \item [Day 7:] For today, try to learn some \textbf{practical tips and strategies to continue eating healthy meals while traveling or on a budget}. Sticking to a nutritious meal plan can feel challenging when one is away from home or trying to save money, but it is doable with the right approach. Feel free to explore strategies like making smart choices when eating out, drinking enough water, planning simple, nutritious meals in advance, or shopping wisely. 

    \item [Day 8:] As we conclude, take some time today to \textbf{learn about how to set realistic, achievable goals for planning nutritious, healthy meals}. Learn about approaches such as using SMART (specific, measurable, achievable, relevant, time-bound) goals, creating a sustainable, long-term routine, and tracking small wins to stay motivated. 

    We hope that you found the everyday learning tasks to be fruitful, and that you are taking away some practical insights, tools, and skills to make healthy food choices. While our learning plan did not cover everything, we hope it gave you a strong foundation to continue exploring topics in nutrition, meal planning, and healthy living on your own beyond this study. Thank you for your time and participation! 
\end{enumerate}

\section{Knowledge Test} \label{sec:learning_questionnaire}

\begin{enumerate}
    \item What are the three macronutrients? \textit{(used as an attention check)}
            \begin{itemize}
                \item Vitamins, Nutrients, and Omega 3 
                \item Water, Fat and Carbohydrates 
                \item Protein, Fat, and Carbohydrates 
                \item I don't know the answer 
            \end{itemize}

    \item How much protein should an active athlete looking to build muscle consume on average? 
            \begin{itemize}
                \item 0.2 grams per pound of body weight 
                \item 0-0.3 grams per pound of body weight
                \item 0.8-1 grams per pound of body weight
                \item 5 grams per pound of body weight
                \item I don't know the answer
            \end{itemize}

    \item Which of the following is not a benefit of a high-protein diet? 
            \begin{itemize}
                \item Helps maintain and preserve muscle mass 
                \item Improved satiety and reduced hunger 
                \item Significant and rapid weight loss
                \item Aids in recovery from training and physical activity
                \item I don't know the answer
            \end{itemize}

    \item When reading a nutrition label, what is the most important thing to check first? \textit{(used for pre-learning knowledge test as well)}
            \begin{itemize}
                \item Serving size
                \item Total calories
                \item Ingredient list
                \item All the above
                \item I don't know the answer
            \end{itemize}

    \item Which of the following describes a practical approach to shopping for groceries? \textit{(used for pre-learning knowledge test as well)}
            \begin{itemize}
                \item Choose foods from the perimeter of the grocery store, as items stocked here tend to be less processed than other foods.
                \item Prioritize organic foods (like an organic candy bar) even if they are more highly processed than non-organic foods (like a non-organic bunch of spinach).
                \item Focus on purchasing foods stocked at eye-level on grocery store shelves, which are almost always less processed than other foods.
                \item Choose foods with more ingredients, particularly if they include added vitamins and minerals.
                \item Shopping while hungry tends to increase the purchasing of healthier, less processed foods. 
            \end{itemize}

    \item How many calories are in 1 gram of carbohydrate, fat, or protein? 
            \begin{itemize}
                \item 4, 9, 4
                \item 3, 9, 6
                \item 7, 9, 4
                \item 5, 5, 9
                \item I don't know the answer
            \end{itemize} 

    \item In considering animal and plant sources of protein, which of the following is true? 
            \begin{itemize}
                \item Animal sources of protein tend to be incomplete because they do not provide all the essential amino acids in adequate amounts to be considered complete proteins. 
                \item The human body cannot make any amino acids, so we need to get these from protein-rich foods. 
                \item In general, animal sources of protein are overall healthier sources of dietary fiber than plant sources of protein.
                \item In many global food traditions, two or more plant sources of protein are combined to enhance the amino acid profile of a dish, incorporating more essential amino acids into the diet.
                \item In general, meat-based diets tend to be lower in saturated fat than vegetarian diets.
            \end{itemize}

    \item Read the following nutrition label for a flavored yogurt. Based on the label, which of the following interpretations is the most accurate? \textit{(used for pre-learning knowledge test as well)}
            \begin{itemize}
                \item The product contains 18 g of sugar from fruit and dairy alone.
                \item Only the added sugars contribute to total caloric intake, naturally occurring sugars are not included in the calorie counts.
                \item The yogurt contains 10 g of sugar added during manufacturing, while the remaining 8 g are intrinsic to ingredients.
                \item Because total carbohydrate is 22 g and total sugar is 18 g, the other 4 g must be added sugars not disclosed in the label.
                \item I don't know the answer.
            \end{itemize}

    \item Which of the following is the most impacted by batch-cooking and reheating? 
            \begin{itemize}
                \item Simple sugars. 
                \item Unsaturated fats.
                \item Water-soluble vitamins.
                \item Dietary fiber.
                \item I don't know the answer.
            \end{itemize}

    \item Which of the following statements is true? 
            \begin{itemize}
                \item Eating most of your daily calories at night enhance insulin sensitivity.
                \item Timing has no effect on macronutrient metabolism, only total intake matters.
                \item Protein oxidation peaks overnight, making late-night protein optimal.
                \item Carbohydrates consumed in the evening lead to higher glucose or insulin responses than identical morning servings.
                \item I don't know the answer.
            \end{itemize}

    \item Compare and contrast the following two meal plans. What are the potential nutritional trade-offs someone might face when choosing one meal plan over the other? Briefly explain your reasoning. 

    \item Compare nutrition labels for the following two granola bars. Which one would be suited for hikers? Briefly explain your answer.

    \item Consider the lunch meal plan below. What changes will you make to improve its nutritional quality, while keeping it enjoyable, satisfying, and budget-friendly? Explain your reasoning. 

    \item We would like you to create a 3-day meal plan for breakfast based on your learning/preferences. Share a couple sentences describing why you chose each item.

\end{enumerate}

\section{Rubric for Open-ended Questions}\label{sec:rubric}

\subsection{Q11: Comparing two Meal Plans}
In Q11, participants were asked to compare two meal plans, and comment on potential nutritional trade-offs when selecting one over the other. We followed a binary 0 (correct), 1 (incorrect) rubric, where responses were penalized for providing (1) incorrect reasoning and (2) information irrelevant to the question asked.

\subsection{Q12: Comparing two Nutrition Labels}
As part of Q12, participants were asked to critically evaluate the nutrition labels of two granola bars and select the one more suitable for hikers. We followed a binary 0 (correct), 1 (incorrect) rubric, where responses were penalized for (1) incorrect interpretation of the nutrition labels and (2) providing incorrect, vague, or generic reasoning. 

\subsection{Q13: Improving a Lunch Meal Plan} 
In Q13, participants were asked to improve the nutritional quality of a given lunch meal plan, while keeping it enjoyable and budget-friendly. After reviewing participants' responses, all co-authors finalized the following 1--3 scale rubric for evaluation: 
\begin{enumerate}
    \item Poor: Makes no or minimal adjustments to improve the meal’s nutritional quality. Provides little or no reasoning. Example response: ``I would keep it as is.''
    
    \item Good: Suggests one strong change that improves the nutritional quality of the meal plan. Reasoning may be brief or limited. May not explicitly consider improvements that are enjoyable and budget-friendly. Example response: ``I may seek to increase the protein in the meal by adding something like beans, lentils or grilled chicken. These are all simple ingredients that can be purchased at a low cost and add nutritional value to this meal.''

    \item Excellent: Suggests multiple changes that improve the nutritional quality of the meal plan. Provides clear, well-structured reasoning demonstrating an understanding of nutritional trade-offs and explicitly considers both enjoyment and budget-friendliness. Example response: ``Brown rice with grilled chicken, more protein. Greens and olive oil would stay, though I would consider adding a different dressing for taste sake. To remove the added sugars, I would get rid of the fruit yogurt, and add plain yogurt, with some honey and maybe some fresh fruit.''
\end{enumerate}

\subsection{Q14: Creating a 3-day Meal Plan}
In Q14, participants were asked to create a 3-day meal plan for breakfast based on their learning. After going through a subset of participants' responses, and considering existing guidelines on meal planning for healthy eating~\cite{mayoclinicHealthyMeals,usda_guidelines} we rated each response on a 1--3 discrete scale (representing low, moderate, and high alignment), on three dimensions: (1) Personalization; \textit{meal plan is customized to personal preferences such as, meal prep time, budget constraints, fitness goals, or dietary restrictions}, (2) Variety; \textit{meal plan provides varied recommendations for each day}, and (3) Balanced, nutrient-dense intake; \textit{meal plan focuses on meeting food group needs with nutrient-dense choices}. Finally, ratings across the three dimensions were averaged to generate the final score. 

\section{Screening Survey} \label{sec:screening_survey}

\begin{enumerate}
    \item How often do you use generative AI tools or systems (e.g., ChatGPT, Bing Copilot, Gemini, etc.)? 
        \begin{itemize}
            \item Never 
            \item Occasionally, about a couple of times a month 
            \item Sometimes, about a couple of times a week 
            \item Often, about once a day 
            \item All the time, or many times a day 
        \end{itemize}

    If you are a good fit for the study, we will ask you to learn about ``nutrition and meal planning'' over the next few days. To help us understand your current knowledge and interest in this topic, please answer the following questions: 

    \item How would you rate your overall knowledge about nutrition and meal planning? 
        \begin{itemize}
            \item Very poor, I do not know anything about meal planning
            \item Poor, I know very little about nutrition and meal planning 
            \item Fair, I know some basics of nutrition and meal planning 
            \item Good, I have a decent knowledge of nutrition and meal planning 
            \item Excellent, I know a lot about nutrition and meal planning
        \end{itemize}

    \item How would you rate your overall interest in nutrition and meal planning or learning about nutrition and meal planning? 
        \begin{itemize}
            \item Very low, not at all interested 
            \item Low, slightly interested 
            \item Moderate, somewhat interested 
            \item High, very interested 
            \item Very high, extremely interested 
        \end{itemize}

    \item We may ask you to look up information about topics such as weight loss, dieting, or healthy eating. Would you feel comfortable doing so? 
        \begin{itemize}
            \item Yes 
            \item No 
        \end{itemize}

    \item How often do you engage in learning new things on your own, outside of work or classroom settings (e.g., by looking up online, reading books, or following social media)? 
        \begin{itemize}
                \item Never 
                \item Occasionally, about a couple of times a month 
                \item Sometimes, about a couple of times a week 
                \item Often, about once a day 
                \item All the time, or many times a day 
        \end{itemize}

    \item If you are a good fit, you will be asked to participate in a 10-day long study (excluding weekends). You will be required to complete daily tasks, including writing diary reflections. Would you be willing to participate? 
        \begin{itemize}
                \item Yes; if selected, I will participate regularly over the 10-day period 
                \item No, thanks 
        \end{itemize}
\end{enumerate}

\section{Onboarding Survey} \label{sec:onboarding_survey}

\subsection{Consent}
\begin{enumerate}
    \item Would you like to participate in this study, as described in the consent form?  
        \begin{itemize}
            \item Yes, I would like to participate 
            \item No, thanks
        \end{itemize}
        
    \item Do you agree to being audio/video recorded if you are invited to participate in the interview study and choose to accept? 
        \begin{itemize}
            \item Yes
            \item No
        \end{itemize}
\end{enumerate}

\subsection{Demographic Information}
\begin{enumerate}
    \item What is your age? 
        \begin{itemize}
            \item 18-25 
            \item 26-35 
            \item 36-45 
            \item 46-55 
            \item >55 
            \item Prefer to not say 
        \end{itemize}
    \item What is your gender? 
        \begin{itemize}
            \item Man 
            \item Woman 
            \item Non-binary 
            \item Self-described: \rule{1cm}{0.15mm}
            \item Prefer to not say 
        \end{itemize}
    \item What is your highest completed level of education? 
        \begin{itemize}
            \item High school 
            \item University or College
            \item Graduate degree
            \item Other: \rule{1cm}{0.15mm}
        \end{itemize}
\end{enumerate}

\subsection{Pre-learning Knowledge Test}
Please answer the following questions based on what you already know about nutrition and meal planning. Answer them with your best knowledge without consulting other people or tools. Incorrect answers will not impact your participation or compensation. 

\begin{enumerate}
    \item When reading a nutrition label, what is the most important thing to check first?
            \begin{itemize}
                \item Serving size
                \item Total calories
                \item Ingredient list
                \item All the above
                \item I don't know the answer
            \end{itemize}

    \item Which of the following describes a practical approach to shopping for groceries?
            \begin{itemize}
                \item Choose foods from the perimeter of the grocery store, as items stocked here tend to be less processed than other foods.
                \item Prioritize organic foods (like an organic candy bar) even if they are more highly processed than non-organic foods (like a non-organic bunch of spinach).
                \item Focus on purchasing foods stocked at eye-level on grocery store shelves, which are almost always less processed than other foods.
                \item Choose foods with more ingredients, particularly if they include added vitamins and minerals.
                \item Shopping while hungry tends to increase the purchasing of healthier, less processed foods. 
            \end{itemize}

    \item Read the following nutrition label for a flavored yogurt. Based on the label, which of the following interpretations is the most accurate?
            \begin{itemize}
                \item The product contains 18 g of sugar from fruit and dairy alone.
                \item Only the added sugars contribute to total caloric intake, naturally occurring sugars are not included in the calorie counts.
                \item The yogurt contains 10 g of sugar added during manufacturing, while the remaining 8 g are intrinsic to ingredients.
                \item Because total carbohydrate is 22 g and total sugar is 18 g, the other 4 g must be added sugars not disclosed in the label.
                \item I don't know the answer.
            \end{itemize}
\end{enumerate}

\subsection{Onboarding for the daily tasks }
 
The main part of the study consists of 8 days of daily tasks. Each day, starting tomorrow, you will be invited to a new Prolific study. The study will include: 

\begin{itemize}
    \item Link to a survey page where you will be introduced to the learning topic for the day and a set of questions for confirming completion of the daily learning task.
    \item After you submit the above survey, you will be redirected to a separate link to submit a diary reflecting on your learning experience and other information-seeking activities you engaged in.
\end{itemize}
 
\noindent\textbf{Learning Goal: }

\noindent In this study, you will be asked to learn about nutrition and meal planning. This is a topic that interests many, with the goal of being able to design healthy, nutritious, or budget-friendly meals for oneself or one’s family based on their needs (e.g., to be healthier, lose weight, or build muscles), or even becoming a certified nutritionist or meal planner to work with clients. Each day, we will suggest some sub-topics to help you structure your self-directed learning process. We hope you find these daily topics interesting, and that the learning is valuable to you beyond the duration of the study.

\noindent\textbf{Assigned technology:  }

\noindent To learn about our provided topics your assigned technology is [ChatGPT/Google]. Please follow the steps below to setup this technology: 

\noindent [Instructions for ChatGPT group]: 

\begin{itemize}
    \item Use this link to access ChatGPT: https://chatgpt.com/  
    \item You do not need to sign in or create an account.
\end{itemize}

\noindent [Instructions for Google group]: 

\begin{itemize}
    \item Use this link to access Google: https://www.google.com/ 
    \item You do not need to sign in or create an account.
    \item IMPORTANT: Google recently introduced an AI Overview feature that appears on many search result pages. However, in this study, we would like to avoid the impact of AI-generated results. Hence, for the duration of this study, we ask you to add ``-ai'' to all your search queries when using Google. For example, if you want to look up the definition of nutrition, you may use: ``what is nutrition -ai'' or ``define nutrition -ai''. 
\end{itemize} 

\noindent Take some time to answer the following questions. These will help us confirm that your assigned technology is set up correctly and that you will be able to use it consistently throughout the study duration. 

\begin{itemize}
    \item Do you agree to use [ChatGPT/Google] as your primary information-seeking channel throughout the study period? This means that you will use it to complete the assigned learning tasks. For your other information needs you are encouraged to go to this technology first, but feel free to use your usual information-seeking channels if needed. 
        \begin{itemize}
            \item Yes, I agree 
            \item No, thanks
        \end{itemize}

    \item Using the assigned technology, please perform a simple information-seeking task:  look up the definition of meal planning. Take a screenshot of [your interaction with ChatGPT/the Google search results page] (with a full screen view to help us see what you did) and upload it here. 
\end{itemize}

\noindent \textbf{Diary entry practice: }

\noindent Since a big part of this study involves writing diary reflections, we would like you to try writing one now. This will give us a chance to review your submission and provide helpful feedback through Prolific messages, if needed. 

\noindent Write a diary entry reflecting on the last time you sought information (e.g., to learn about a topic or to answer questions). Describe your information need (for what topic or question you needed to seek information), how you tried to satisfy this need (e.g., what sources, tools, or strategies you used, and why you chose them), what the outcome was, and how you felt about this information-seeking experience. Describe any parts of the information-seeking process that felt unsatisfying or challenging, and how you addressed them.

\end{document}